# Normalized Blood Flow Index in Optical Coherence Tomography Angiography Provides a Sensitive Biomarker of Early Diabetic Retinopathy

Albert K. Dadzie[1]; David Le[1]; Mansour Abtahi[1]; Behrouz Ebrahimi[1]; Taeyoon Son[1]; Jennifer I. Lim[2]; Xincheng Yao[1,2]

[1]Department of Biomedical Engineering, University of Illinois at Chicago, Chicago, IL 60607, USA, [2]Department of Ophthalmology and Visual Sciences, University of Illinois at Chicago, Chicago, IL, 60612, USA

**Corresponding author:**

Xincheng Yao, PhD

Richard & Loan Hill Professor

Department of Biomedical Engineering (MC 563)

University of Illinois at Chicago

Clinical Sciences North, Suite W103, Room 164D

820 South Wood Street, Chicago, IL 60612

Tel: (312)413-2016; Fax: (312)996-4644; Email: xcy@uic.edu

**Word Count:** 2763

**Funding/Support:** National Eye Institute (R01 EY023522, R01 EY029673, R01 EY030101, R01 EY030842, P30EY001792); Research to Prevent Blindness; Richard and Loan Hill Endowment.

**Commercial Relationships:** None

**Abstract**

**Purpose:** To evaluate the sensitivity of normalized blood flow index (NBFI) for detecting early diabetic retinopathy (DR).

**Methods:** Optical coherence tomography angiography (OCTA) images of 30 eyes from 20 healthy controls, 21 eyes of diabetic patients with no DR (NoDR) and 26 eyes from 22 patients with mild non-proliferative DR (NPDR) were analyzed in this study. The OCTA images were centered on the fovea and covered a 6 mm x 6 mm area. Enface projections of the superficial vascular plexus (SVP) and the deep capillary plexus (DCP) were obtained for the quantitative OCTA feature analysis. Three quantitative OCTA features were examined: blood vessel density (BVD), blood flow flux (BFF), and normalized blood flow index (NBFI). Each feature was calculated from both the SVP and DCP and their sensitivity to distinguish the three cohorts of the study were evaluated.

**Results:** The only quantitative feature that was capable of distinguishing between all three cohorts was NBFI in the DCP image. Comparative study revealed that both BVD and BFF were able to distinguish the controls from NoDR and mild NPDR. However, neither BVD nor BFF was sensitive enough to separate NoDR from the healthy controls.

**Conclusion:** The NBFI has been demonstrated as a sensitive biomarker of early DR, revealing retinal blood flow abnormality better than traditional BVD and BFF. The NBFI in the DCP was verified as the most sensitive biomarker, supporting that diabetes affects the DCP earlier than SVP in DR.

**Introduction**

Diabetes is one of the world's most common noncommunicable diseases, as well as one of the top causes of death in developed countries[1]. Diabetic retinopathy (DR) is the most common complication of diabetes mellitus and is the leading cause of vision loss in developed countries[2]. The number of people with diabetes was 285 million in 2010 and is estimated to grow about 69% by 2030[3]. In United States, it is estimated that about 28.5% of diabetic adults have DR. The prevalence of DR is expected to rise significantly as the prevalence of diabetes is increasing exponentially[4]. Most of these patients that develop DR are unaware of the early retinal abnormalities. This is because the early stages are typically asymptomatic until the late stages, when damage is irreversible[5,6]. Therefore, if irreversible damage to the eye is to be prevented, early diagnosis of the disease is of great importance. As such, accurate investigative techniques are becoming increasingly vital in the early detection and diagnosis of DR for effective management of the disease. Optical coherence tomography angiography (OCTA) is a new technology that is being extensively studied and it is showing great potential for application for early detection of DR.

As a modality extension of optical coherence tomography (OCT), OCTA provides a noninvasive imaging technology that can be used to generate angiographic images from blood flow information[7], without the involvement of exogenous labels required in traditional fluorescein angiography (FA). This revolutionary technology is characterized by its high imaging resolution at microcapillary level. It has been demonstrated that the OCTA increases the visibility of all vascular layers including layers that are not well-distinguished on FA[8]. This indicates that OCTA can be used to effectively diagnose microvascular diseases like DR. Quantitative OCTA features, such as blood vessel caliber (BVC), blood vessel tortuosity (BVT) and fovea avascular zone (FAZ), have been extensively explored for DR detection and staging classification[9-14].

One of the most studied quantitative OCTA features is blood vessel density (BVD) also referred to as vessel area density (VAD), vessel density (VD) or capillary density. It has been studied as a biomarker for several diseases including DR[9-13], age-related macular degeneration[15,16], glaucoma[17,18] and sickle cell retinopathy[19,20]. Conventionally, BVD has been used to indicate blood flow within the retina. However, a recent study by Abdolahi et al., explained that BVD only provides limited information on blood flow because it simply detects the presence or absence of blood flow[21]. Previously, the intensity of the optical microangiography (OMAG) signal was shown to be related to the number of red blood cells (RBCs) flowing within the vessels[22]. This phenomenon is termed blood flow flux (BFF) and it approximates the number of RBCs flowing through capillary segments and it has been shown to be a more robust measure for assessing subclinical blood flow than BVD[21,23]. In the calculation of BFF, the absolute (non-binarized) decorrelation intensity values from the OCTA image are averaged. A potential complication of quantitative BFF analysis is that inevitable variabilities, such as subject pigmentation level, eye condition, ocular transparency, illumination irradiance, and detector sensitivity, may affect the OCTA signal magnitude, which will impact differential BFF analysis among patients. Moreover, different image processing algorithms used to generate the OCTA images, voluntary and involuntary eye movements, and blinking cause inevitable noises to further complicate OCTA signal quantification[24]. In this study, we propose a normalized blood flow index (NBFI) to compensate for potential variabilities and noises. This NBFI was employed as a quantitative feature to differentiate normal eyes, NoDR and mild NPDR.

**Methods**

In this retrospective study, we assessed NBFI as a sensitive biomarker for the early detection of DR. The study was approved by the Institutional Review Board (IRB) of the University of Illinois at Chicago (UIC) and was in accordance with the guidelines in the Declaration of Helsinki. The participants in this study were recruited from the UIC retinal clinic. OCTA images were obtained from 77 eyes and grouped into three categories. The first group was made of 30 images obtained from healthy participants with no retinopathies. The second group was made up of 21 images obtained from NoDR patients. The third group which included 26 images obtained from mild NPDR patients according to the Early Treatment Diabetic Retinopathy Study (ETDRS) staging system. Only individuals who were 18 years or older were included in this study. Participants with macular edema, prior vitrectomy surgery, opaque ocular media, age-related macular degeneration, or any other retinal diseases were excluded from this study. All patients underwent a comprehensive ophthalmic examination that included detailed history taking, best corrected visual acuity (BCVA), detailed anterior segment examination using slit-lamp examination, dilated fundus examination using both biomicroscope and indirect ophthalmoscopy. A single retina specialist classified the patients as having no DR (NoDR) or having mild NPDR using the ETDRS staging system.

The ANGIOVUE spectral domain OCT (Optovue, Fremont, CA) equipment was used to obtain 6 mm x 6 mm macular scans for each subject. All images were manually evaluated and only images with signal quality >6 and minimal motion artifact were included in the study. The OCTA enface images of both the SVP and DCP were exported using the built-in ReVue program in the ANGIOVUE system. The OCTA images were imported into MATLAB (MathWorks, Natick, MA) software that was custom developed for quantitative feature extraction and analysis.

In this study, three quantitative OCTA features were analyzed, i.e., BVD, BFF and NBFI. For the evaluation of BFF, the mean of the pixel intensities of the non-binarized OCTA image (Fig 1A) was calculated. A vessel map was generated by using a thresholding algorithm to the OCTA image to create a binary image that assigned a value of 1 to pixels that corresponded to perfusion and a value of 0 to areas that correspond to background (Fig 1B). From this vessel map, BVD was calculated as the ratio of the number of pixels that represent perfusion to the total number of pixels in the OCTA image. The perfusion map (Fig 1C) was generated by using the vessel map as a mask to select the portions of the OCTA image that corresponds to perfusion. A detailed explanation of this is explained by Yao et al.[14] The noise map (Fig 2D) was generated by using the inverse of the vessel map as a mask to select portions of the OCTA image not covered by blood vessels. The NBFI was then calculated by dividing the mean of the perfusion map by the standard deviation of the noise map. Figure 2 shows the representative OCTA images, perfusion maps and noise maps of both the SVP and DCP of controls, NoDR and mild NPDR.

Statistical analysis was performed using R Software, version 4.2.0 (R Core Team, Vienna, Austria). The Shapiro-Wilk test was first used to check for normality for all the quantitative features. Multiple group comparisons for the normally distributed features were conducted using the one-way ANOVA test, followed by individual pairwise comparisons using the unpaired Student's t-test. Conversely, the Kruskal-Wallis one-way ANOVA test was used for multiple group comparison for features that were not normally distributed. This was followed by individual pairwise comparison using the Mann-Whitney t-test. A p value of <0.05 was considered statistically significant in this study.

**Results**

A total of 30 eyes from 20 healthy controls, 21 eyes from 15 NoDR patients and 26 eyes from 22 patients with mild NPDR were imaged for this study. A summary of the characteristics of the participants and their eyes are shown in Table 1. There were no significant differences in age, sex, or history of hypertension between the controls and the diabetic patients (ANOVA, P = 0.69, chi-square test, P = 0.85). Also, no significant differences were observed in the duration of diabetes or insulin dependency among the diabetic groups.

Comparative analysis of the quantitative features is summarized in Table 2. In both the SVP and DCP, BVD showed a decrease from the normal to the NoDR group and then increased in the mild NPDR group. In the SVP, BVD was sensitive to distinguish between eyes with no retinopathy (normal + NoDR) and eyes with Mild NPDR. In the DCP however, BVD only showed significant difference between the diabetics (i.e., NoDR vs Mild NPDR, P = 0.04). The other comparisons lacked statistical significance. Analysis of BFF showed a decreasing trend from the normal to the NoDR group and further decrease in the mild NPDR group in both the SVP and DCP. The results suggested that BFF was not a sensitive feature in the SVP. It was unable to statistically distinguish between the three groups studied (P>0.05). However, in the DCP, it could significantly distinguish between eyes with no retinopathy (normal + NoDR) and eyes with mild NPDR. On the other hand, NBFI increased among patients with mild NPDR compared to both the controls and NoDR groups in both the SVP and DCP. The NBFI in both layers showed good significance but only DCP was sensitive enough to distinguish between all the three groups. In the SVP, NBFI was only sensitive enough to separate eyes with no retinopathies (control + NoDR) and eyes with mild NPDR.

**Discussion**

Quantitative features such as BVD[12,13], BVC[9], BVT[25], and FAZ area[11] have been developed and validated for DR diagnosis and staging. The most popular vascular based quantitative feature by far is BVD. By determining the percentage of the OCTA image that is occupied by blood vessels, BVD shows how much perfusion is present. Several studies have evaluated this feature as a sensitive biomarker to DR detection and staging[9-13,26]. Although many of these studies concur on the sensitivity of BVD, there are some disagreements regarding how the various stages of the disease process affect the quantification of this feature. Most researchers agree that capillary dropout causes BVD to decline as the disease progresses[9-13]. Meanwhile, other studies such as Mastropasqua et al. report an initial increase in BVD in the early stages before a decline is noted[11]. A similar trend was observed in this study. The presence of microaneurysms which is typical of mild NPDR coupled with vessel dilation which has been shown to occur in DR[9,27], could possibly be the cause of BVD inflation in the mild NPDR group. This means that the increase in BVD noted in the mild NPDR group does not necessarily imply increased vascular structures. Despite variations in how BVD is affected by the severity of DR, all these studies, including our study, agree that BVD can be a useful biomarker for the diagnosis of the disease. However, the sensitivity of BVD can only be appreciated in the later stages of the disease. From this study, BVD was not sensitive enough to discriminate between normal healthy eyes and diabetic eyes which have not yet developed any retinopathy. This is consistent with findings from previous studies[11,28]. Therefore, for early detection of DR, a more sensitive feature needs to be explored. Moreover, recent studies criticize the appropriateness of using BVD to describe changes in blood flow in DR[21,23,29]. They suggest that BFF, as opposed to BVD, may be a more precise and sensitive indicator of retinal perfusion, particularly when it comes to subclinical alterations that take place in the early stages of DR. This is because BFF, which estimates the average number of red blood cells (RBCs) moving in the vessels, is basically the same as estimating the amount of blood flow in the vessels[22].

As a relatively new quantitative OCTA feature, BFF accurately quantifies blood flow in retinal vessels from OCTA images. Recent studies have attempted to explain changes in blood flow as a result of diabetes using BFF. A study by Choi et al. demonstrated using OMAG that the signal intensity (or strength) of a particular voxel is proportional to the number of particles across the imaging voxel cross-section per unit time[22]. This implies that the concentration of particles (RBCs) directly relates to the intensity of the pixels in OCTA images. Therefore, BFF is calculated by finding the average of the pixel intensities of a grayscale OCTA image. Some studies have been conducted to verify the importance of this feature and they conclude that BFF provides more information about retinal perfusion and is more sensitive to subclinical changes in retinal blood flow[21,23]. Also, BFF has been shown to be significantly correlated with hematocrit and hemoglobin concentration, showing that it is a superior indicator of blood flow than BVD[23]. An interesting observation made by these studies, however, is that although BFF gives a more precise quantification of subclinical changes in blood flow, it was not found to be significantly correlated with the presence of retinopathy. A possible explanation for this reduced sensitivity could be the presence of intrinsic noise found in OCTA images that might affect the calculation of BFF. Noise is an inevitable component of OCTA images and are caused by a myriad of factors including but not limited to the image processing technique used to generate the images, voluntary and involuntary eye movements and blinking[24]. To counter the effect of noise in the calculation of BFF, a normalization technique was introduced in this study to reduce the effect of noise and also to improve the sensitivity of BFF for early detection of DR. This new feature, we termed it as NBFI.

As a new OCTA feature, NBFI was observed to be more sensitive than conventional BFF to differentiate the early stages of DR. Previous studies, including this current study show that BFF progressively decrease as the severity of the disease increases, indicating a reduction in blood flow with increasing severity of DR[21,23]. Meanwhile, NBFI did not follow this trend. In both the SVP and DCP, NBFI in the mild NPDR group was significantly higher than both the control and NoDR groups. This finding suggests that, in the very early stages of DR, blood flow increases initially before declining. The same observations were reported in studies that examined blood flow in the retina of diabetic patients. Using laser velocimetry, Patel et al., as early as 1992, observed that blood flow significantly increased in subjects with DR compared to those without DR (control + NoDR) and subsequently began to decline as the severity of DR increased[30]. In a recent study by Ueno et al. in 2021, they used laser speckle flowgraphy (LSFG) to confirm that the blood flow in patients with mild and moderate NPDR is higher than those without retinopathy[31]. The trend observed by these studies were similar to the trend in our study, indicating that NBFI might be a better reflection of blood flow than conventional BFF. It has already been established that capillary dropout occurs even in the early stages of the disease[9,11,13] and therefore the observed increase in blood flow could be explained to be a compensatory mechanism to maintain adequate perfusion in order to meet the metabolic demands of the retina.

Retinal blood flow is a critical property that can be used in the detection, diagnosis, and monitoring of DR. A variety of OCTA quantitative features have been proposed and evaluated to give an insight into how DR is affects blood flow in the retina. According to our findings, NBFI is the feature that most accurately depicts retinal blood flow. Also, NBFI showed the highest sensitivity in discriminating between the early stages of DR. The NBFI in the DCP was the only feature that was sensitive enough to differentiate between all the three groups that were studied. This suggests that in the very early stages of DR, subclinical changes in retinal blood flow may occur in the DCP earlier than in the SVP. This is consistent with a review of the pathogenesis of DR where it was concluded that the DCP is affected earlier than SVP in DR[32].

The limitations of this study include the relatively modest sample size and all images were obtained from the same OCTA device. The influence of blood pressure, blood glucose level and duration of diabetes on blood flow was not considered in this study because of the small sample size. Also, different machines may introduce different levels of noise in the OCTA image. Therefore, studies on large samples size using different OCTA devices are necessary to further evaluate the reliability of NBFI as a measure of retinal blood flow in the detection and monitoring of DR.

In conclusion, this study has validated NBFI as a sensitive biomarker of early DR. Comparative analysis indicated that NBFI can reveal retinal blood flow abnormality better than previously proposed BVD and BFF. The NBFI in the DCP was verified as the most sensitive biomarker to detect early DR. This observation supports that diabetes affects the DCP earlier than SVP in the retina.

Table 1.

| TABLE 1. Demographic of the Healthy Subjects, NoDR and Mild NPDR patients. | | | |
|---|---|---|---|
| | Healthy Subjects | NoDR | Mild NPDR |
| No. of subjects (n) | 20 | 15 | 22 |
| Age (years) | 52.6 ± 14.57 | 59.33 ± 11.31 | 62.14 ± 11.99 |
| Age range | 35 – 80 | 40 – 80 | 24 – 78 |
| Gender (male/female) | 12/8 | 5/10 | 8/14 |
| Duration of diabetes (years) | - | 8.89 ± 5.04 | 16.73 ± 4.82 |
| Diabetes type | - | Type II | Type II |
| Insulin dependence (Y/N) | - | 3/11 | 5/12 |
| HbA1c, (%) | - | 7.6 ± 2.1 | 8.3 ± 2.4 |
| HTN prevalence, % | 10 | 72 | 70 |
| HbA1C = glycated hemoglobin; HTN = hypertension. | | | |

Table 2.

| TABLE 2. Comparisons of OCTA quantitative features between controls, NoDR and mild NPDR groups | | | | | | | |
|---|---|---|---|---|---|---|---|
| | | | | p value | | | |
| Feature | Control (I) | No DR (II) | Mild (III) | I vs II | I vs III | II vs III | ANOVA |
| BVD sup | 0.4315 ± 0.0178 | 0.4307 ± 0.0183 | 0.4558 ± 0.0220 | 0.8775 | <.0001 | 0.0001 | <.0001* |
| BVD deep | 0.4360 ± 0.0343 | 0.4146 ± 0.0181 | 0.4438 ± 0.0311 | 0.0564 | 0.1988 | 0.004 | 0.0042† |
| BFF sup | 0.2215 ± 0.0226 | 0.2199 ± 0.0304 | 0.2121 ± 0.0226 | 0.8374 | 0.1239 | 0.3315 | 0.3391* |
| BFF deep | 0.2422 ± 0.0303 | 0.2375 ± 0.0444 | 0.2073 ± 0.0331 | 0.6765 | 0.0002 | 0.0140 | 0.0011* |
| NBFI sup | 2.8827 ± 0.3004 | 2.9216 ± 0.4297 | 3.2907 ± 0.3961 | 0.8570 | <.0001 | 0.0002 | <.0001† |
| NBFI deep | 2.8405 ± 0.4187 | 2.6387 ± 0.2694 | 3.1254 ± 0.3718 | 0.0416 | 0.0093 | <.0001 | <.0001* |

BVD = blood vessel density, BFF = blood flow flux, NBFI = normalized blood flow index, sup = superficial
*Multiple group comparisons performed using one-way ANOVA, corresponding individual comparisons were performed using Student's t-test. †Multiple group comparisons performed using Kruskal–Wallis one-way ANOVA, corresponding individual comparisons were performed using Mann-Whitney's t-test. Statistical significance p value <0.05.
Data are expressed as mean ± SD.

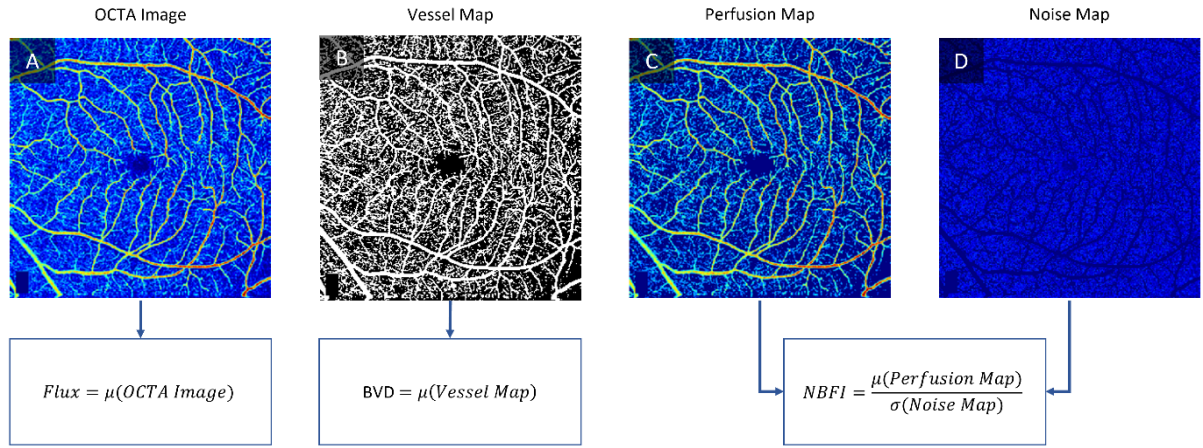

**Figure 1.** Image processing steps in quantification of OCTA features. (A) Representative OCTA image. (B) Binarized map of the OCTA image used for quantifying BVD. The binarized image was then used as a mask to extract areas that correspond to perfusion and areas that correspond to noise separately. (C) OCTA image with the noise removed. (D) Background noise removed from the OCTA image. From C and D, NBFI was calculated.

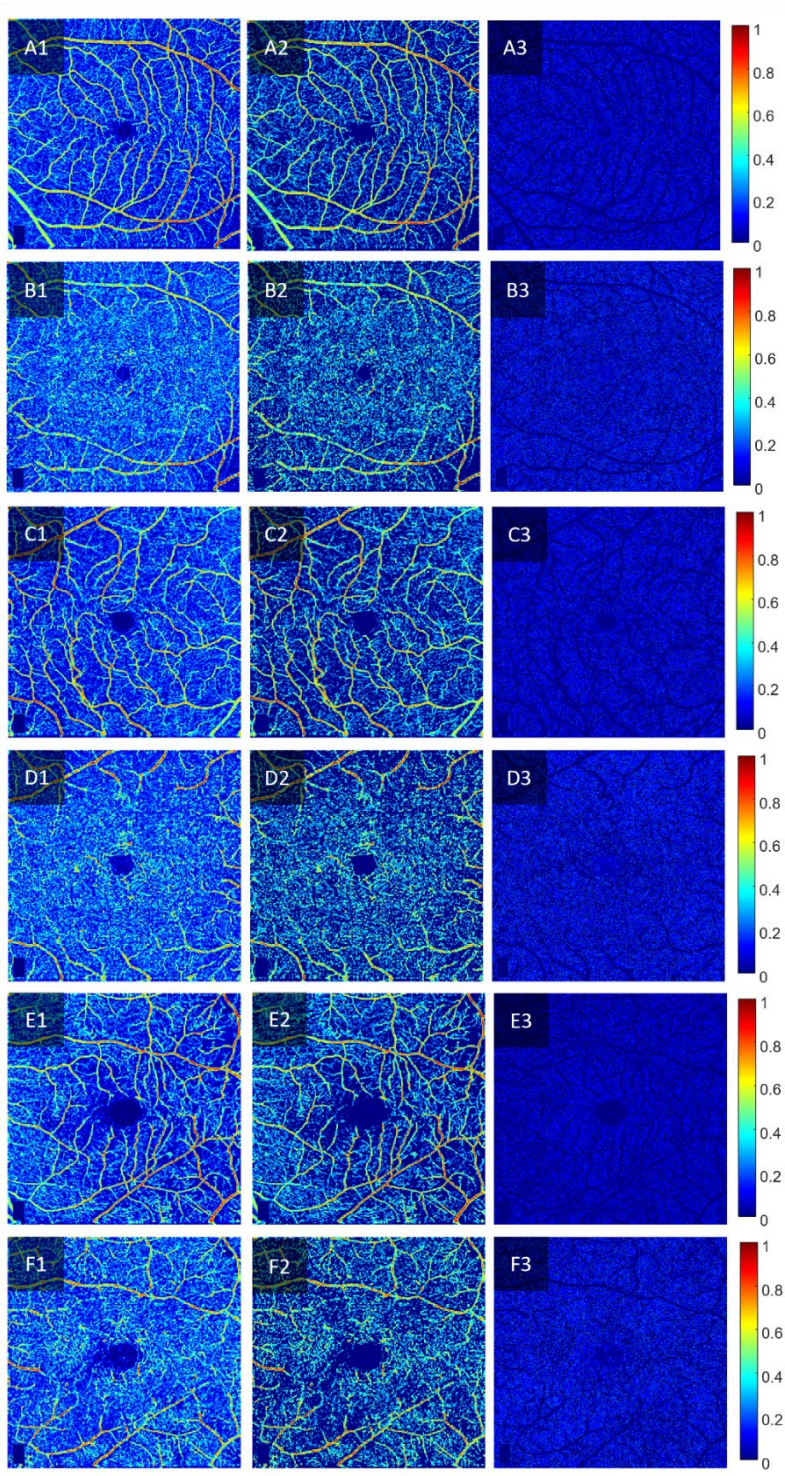

**Figure 2.** Representative OCTA images of the SVP of controls (A1-A3), the DCP of controls (B1-B3), SVP of NoDR (C1-C3), DCP of NoDR (D1-D3), SVP of mild NPDR (E1-E3) and DCP of mild NPDR (F1-F3). Column 1 shows the pseudo color flux map, column 2 shows the pseudo color flux map with the noise removed and column 3 shows the pseudo color representation of the noise in the OCTA image.